\documentclass{ws-procs9x6}

\begin{document}

\newcommand{\quarter}{{\textstyle\frac{1}{4}}}
\newcommand{\const}{\mathit{const}}

\newcommand{\fm}{\mathrm{fm}}
\newcommand{\MeV}{\mathrm{MeV}}
\newcommand{\GeV}{\mathrm{GeV}}

\renewcommand{\vec}[1]{\boldsymbol{{#1}}}
\newcommand{\unitvec}[1]{\hat{\vec{#1}}}

\newcommand{\I}{\mathrm{i}}


\newcommand{\eqdot}{\; .}
\newcommand{\eqcomma}{\; ,}

\newcommand{\bra}[1]{\big< \,{#1}\, \big|}

\newcommand{\ket}[1]{\big| \,{#1}\, \big> }

\newcommand{\braket}[2]{\big< \,{#1}\, \big| \,{#2}\, \big> }
\newcommand{\braketa}[2]{\braket{#1}{#2}{}_a}
\newcommand{\braketaa}[2]{{}_a \braket{#1}{#2}{}_a}

\newcommand{\expect}[1]{\big< \, {#1} \, \big>}

\newcommand{\matrixe}[3]{\big< \,{#1}\, \big| \,{#2}\, 
\big| \,{#3}\, \big> }
\newcommand{\matrixea}[3]{\matrixe{#1}{#2}{#3}{}_a}
\newcommand{\matrixeaa}[3]{{}_a\matrixe{#1}{#2}{#3}{}_a}
\newcommand{\matrixered}[3]{\big<\,{#1}
  \,\big|\big|\,{#2}\,\big|\big|\,{#3}\,\big>}

\newcommand{\comm}[2]{\bigl[ {#1}, {#2} \bigr]_{-} }
\newcommand{\acomm}[2]{\bigl[ {#1}, {#2} \bigr]_{+} }
\newcommand{\commpm}[2]{\bigl[ {#1}, {#2} \bigr]_{\pm} }
\newcommand{\commmp}[2]{\bigl[ {#1}, {#2} \bigr]_{\mp} }

\newcommand{\biggcomm}[2]{\biggl[ {#1}, {#2} \biggr]_{-} }
\newcommand{\biggacomm}[2]{\biggl[ {#1}, {#2} \biggr]_{+} }
\newcommand{\biggcommpm}[2]{\biggl[ {#1}, {#2} \biggr]_{\pm} }
\newcommand{\biggcommmp}[2]{\biggl[ {#1}, {#2} \biggr]_{\mp} }

\newcommand{\partd}[2]{\ensuremath{ \frac{\partial #1}
{\partial #2} }}

\newcommand{\partdd}[2]{\ensuremath{ \frac{\partial^2 #1}
{\partial #2^2} }}

\newcommand{\intd}[1]{\int\!\mathrm{d}{#1}\;}
\newcommand{\intdt}[1]{\int\!\mathrm{d}^3{#1}\;}

\newcommand{\hermit}[1]{{{#1}}^{\dag}}

\newcommand{\op}[1]{{#1}}
\newcommand{\hop}[1]{\hermit{\op{#1}}}

\newcommand{\conop}[1]{\op{#1}^{\dagger}}
\newcommand{\desop}[1]{\op{#1}^{\phantom{\dagger}}}

\newcommand{\opid}{\op{1}}

\newcommand{\VecThree}[3]{%
   \left( \begin{matrix}{#1} \\ {#2} \\ {#3} \end{matrix} \right)}

\newcommand{\Trace}{\mathrm{Tr}}

\newcommand{\grad}[1]{\partd{}{#1}}
\newcommand{\gradx}{\partd{}{\vec{x}}}

\newcommand{\rhoone}{\rho^{(1)}}
\newcommand{\rhotwo}{\rho^{(2)}}
\newcommand{\oprhoone}{\op{\rho}^{(1)}}
\newcommand{\oprhotwo}{\op{\rho}^{(2)}}


\newcommand{\eqrrep}{\overset{\scriptstyle{\vec{r}}}{\Rightarrow}}

\newcommand{\couple}[3]{\left\{ {#1} \: {#2} \right\}^{(#3)}}
\newcommand{\coupletensor}[3]{\left\{ {#1} \otimes {#2} \right\}^{(#3)}}

\newcommand{\couplev}[3]{({#1}\,{#2})^{(#3)}}

\newcommand{\sixj}[6]{\left\{ \begin{matrix} {#1} & {#3} & {#5}\\ {#2}
& {#4} & {#6} \end{matrix} \right\}}
\newcommand{\ninej}[9]{\left\{ \begin{matrix} {#1} & {#4} & {#7}\\ {#2}
& {#5} & {#8} \\ {#3} & {#6} & {#9} \end{matrix} \right\}}
\newcommand{\cg}[6]{\mathrm{C}\Biggl(\,\begin{matrix} {#1} & {#3} \\ {#2}
& {#4} \end{matrix}\, \Biggr|\Biggl.\, \begin{matrix} {#5} \\
{#6} \end{matrix} \,\Biggr)}

\newcommand{\talmi}[4]{\left< \begin{matrix} {#1} \\ {#2} \end{matrix}
    \right|\left. \begin{matrix} {#3} \\ {#4} \end{matrix} \right>}

\newcommand{\corr}[1]{\hat{#1}}

\newcommand{\cop}[1]{\op{\corr{#1}}}

\newcommand{\opC}{\op{C}}
\newcommand{\hopC}{\hop{C}}
\newcommand{\opCr}{\op{C}_{r}^{\phantom{\dagger}}}
\newcommand{\hopCr}{\op{C}_{r}^{\dagger}}
\newcommand{\opCom}{\op{C}_{\Omega}^{\phantom{\dagger}}}
\newcommand{\hopCom}{\op{C}_{\Omega}^{\dagger}}
\newcommand{\opG}{\op{G}}
\newcommand{\opGr}{\op{G}_{r}}
\newcommand{\opGom}{\op{G}_{\Omega}}

\newcommand{\sopLom}{\mathsf{L}_{\Omega}}

\newcommand{\opc}{\op{c}}
\newcommand{\hopc}{\hop{c}}
\newcommand{\opcr}{\op{c}_{r}^{\phantom{\dagger}}}
\newcommand{\hopcr}{\op{c}_{r}^{\dagger}}
\newcommand{\opcom}{\op{c}_{\Omega}^{\phantom{\dagger}}}
\newcommand{\hopcom}{\op{c}_{\Omega}^{\dagger}}
\newcommand{\opg}{\op{g}}
\newcommand{\opgr}{\op{g}_{r}}
\newcommand{\opgom}{\op{g}_{\Omega}}

\newcommand{\opone}[1]{\op{#1}^{[1]}}
\newcommand{\copone}[1]{\cop{#1}^{[1]}}
\newcommand{\optwo}[1]{\op{#1}^{[2]}}
\newcommand{\coptwo}[1]{\cop{#1}^{[2]}}

\newcommand{\icm}{\mathrm{cm}}
\newcommand{\irel}{\mathrm{rel}}
\newcommand{\intr}{\mathrm{intr}}

\newcommand{\vecr}{\vec{r}}
\newcommand{\opvecr}{\op{\vec{r}}}
\newcommand{\vecx}{\vec{x}}
\newcommand{\opvecx}{\op{\vec{x}}}
\newcommand{\vecX}{\vec{X}}
\newcommand{\opvecX}{\op{\vec{X}}}
\newcommand{\vecp}{\vec{p}}
\newcommand{\opvecp}{\op{\vec{p}}}
\newcommand{\vecP}{\vec{P}}
\newcommand{\opvecP}{\op{\vec{P}}}
\newcommand{\vecl}{\vec{l}}
\newcommand{\opvecl}{\op{\vec{l}}}

\newcommand{\vecpr}{\vec{p}_{r}}
\newcommand{\opvecpr}{\op{\vec{p}}_{r}}
\newcommand{\pr}{p_{r}}
\newcommand{\oppr}{\op{p}_{r}}

\newcommand{\pom}{p_{\Omega}}
\newcommand{\vecpom}{\vec{p}_{\Omega}}
\newcommand{\opvecpom}{\op{\vec{p}}_{\Omega}}

\newcommand{\lsq}{\vecl^2}
\newcommand{\oplsq}{\op{\vecl}^2}

\newcommand{\gom}{g_{\Omega}}
\newcommand{\srpom}{s_{\!12}(\vecr,\vecpom)}
\newcommand{\opsrpom}{\op{s}_{\!12}(\vecr,\vecpom)}
\newcommand{\opSrpom}{\op{S}_{\!12}(\vecr,\vecpom)}

\newcommand{\Rp}{R_{+}}
\newcommand{\Rm}{R_{-}}
\newcommand{\Rpm}{R_{\pm}}
\newcommand{\Rmp}{R_{\mp}}
\newcommand{\metricRp}{\mathcal{R}_{+}}
\newcommand{\metricRm}{\mathcal{R}_{-}}
\newcommand{\metricRpm}{\mathcal{R}_{\pm}}
\newcommand{\metricRmp}{\mathcal{R}_{\mp}}

\newcommand{\cmur}{\hat{\mu}_{r}}
\newcommand{\cmuom}{\hat{\mu}_{\Omega}}
\newcommand{\cu}{\hat{u}}

\newcommand{\opSone}{\op{S}^{(1)}}
\newcommand{\opStwo}{\op{S}^{(2)}}

\newcommand{\Pinot}{\Pi_{0}}
\newcommand{\Pione}{\Pi_{1}}
\newcommand{\ls}{\vecl\!\cdot\!\vec{s}}
\newcommand{\lssq}{(\ls)^2}
\newcommand{\srr}{s_{\!12}(\unitvec{r},\unitvec{r})}
\newcommand{\sll}{s_{\!12}(\vecl,\vecl)}
\newcommand{\spompom}{s_{\!12}(\vecpom,\vecpom)}
\newcommand{\sbarpompom}{\bar{s}_{\!12}(\vecpom,\vecpom)}

\newcommand{\opPinot}{\op{\Pi}_{0}}
\newcommand{\opPione}{\op{\Pi}_{1}}
\newcommand{\opls}{\op{\vecl}\!\cdot\!\op{\vec{s}}}
\newcommand{\oplssq}{(\opls)^2}
\newcommand{\opsrr}{\op{s}_{\!12}(\unitvec{r},\unitvec{r})}
\newcommand{\opsll}{\op{s}_{\!12}(\vecl,\vecl)}
\newcommand{\opspompom}{\op{s}_{\!12}(\vecpom,\vecpom)}
\newcommand{\opsbarpompom}{\op{\bar{s}}_{\!12}(\vecpom,\vecpom)}

\newcommand{\opLsq}{\op{\vec{L}}^2}

\newcommand{\lone}{\op{l}^{(1)}}
\newcommand{\rrtwo}{\couplev{\op{\hat{r}}}{\op{\hat{r}}}{2}}
\newcommand{\rpomtwo}{\couplev{\op{r}}{\op{\pom}}{2}}
\newcommand{\lltwo}{\couplev{\op{l}}{\op{l}}{2}}
\newcommand{\pompomtwo}{\couplev{\op{\pom}}{\op{\pom}}{2}}
\newcommand{\pompombartwo}{(\overline{\op{\pom}\,\op{\pom}})^{(2)}}


\newcommand{\crhoone}{\corr{\rho}^{(1)}}
\newcommand{\crhotwo}{\corr{\rho}^{(2)}}

\newcommand{\coprhoone}{\op{\corr{\rho}}^{(1)}}
\newcommand{\coprhotwo}{\op{\corr{\rho}}^{(2)}}

\newcommand{\Vlowk}{\ensuremath{V_{\mathit{low-k}}}}


\title{NUCLEAR STRUCTURE  --  ``ab initio''}

\author{H. FELDMEIER AND T. NEFF}
\address{Gesellschaft f\"ur Schwerionenforschung mbH\\ Planckstr. 1,
         D-64291 Darmstadt, Germany\\ E-mail: h.feldmeier@gsi.de, t.neff@gsi.de}

\author{R. ROTH}
\address{Institut f\"ur Kernphysik\\Schlossgartenstr. 9,
         D-64289 Darmstadt, Germany\\ E-mail: robert.roth@physik.tu-darmstadt.de}


\maketitle

\abstracts{ An ab-initio description of atomic nuclei that solves the
  nuclear many-body problem for realistic nuclear forces is expected
  to possess a high degree of predictive power.  In this contribution
  we treat the main obstacle, namely the short-ranged repulsive and
  tensor correlations induced by the realistic nucleon-nucleon
  interaction, by means of a unitary correlation operator.  This
  correlator applied to uncorrelated many-body states imprints
  short-ranged correlations that cannot be described by product
  states. When applied to an observable it induces the correlations
  into the operator, creating for example a correlated Hamiltonian
  suited for Slater determinants.  Adding to the correlated realistic
  interaction a correction for three-body effects, consisting of a
  momentum-dependent central and spin-orbit two-body potential we
  obtain an effective interaction that is successfully used for all
  nuclei up to mass 60. Various results are shown.}


\section{Introduction}

In the last years exact \emph{ab initio} calculations of light nuclei
have become feasible with Greens Function Monte Carlo
calculations\cite{gfmc} and in the No-Core Shell Model \cite{ncsm}.
Here realistic interactions that fit the nucleon-nucleon scattering
data and the deuteron properties are used \cite{bonn,argonne}.
Additional three-body forces are needed and are adjusted to the
spectra of nuclei. Chiral perturbation promises to provide a
consistent derivation of two- and three-body forces
\cite{entem,gloeckle}.

\section{The Unitary Correlation Operator Method (UCOM)}

Our aim is to perform \emph{ab initio} calculations of larger nuclei
with realistic interactions like the Bonn or Argonne potentials in a
Hartree-Fock picture or a many-body approach with configuration
mixing.

The repulsive core and the strong tensor force of the nuclear
interaction induce strong short-range radial and tensor correlations
in the nuclear many-body system. These correlations are in the
relative coordinates $\vec{r}_{ij}=\vec{r}_i-\vec{r}_j$ and thus can
not be represented by products of single-particle states like Slater
determinants
\begin{equation} \label{Slater}
  \ket{\Psi} = \mathcal{A} \left\{ \ket{q_1} \otimes \ldots \otimes
    \ket{q_A} \right\}
\end{equation}
that are usually used as many-body states in Hartree-Fock or a
shell-model calculations. $\mathcal{A}$ denotes the antisymmetrization
operator and $\ket{q_i}$ the single-particle states.

Instead we treat the radial and tensor correlations explicitly by a
unitary correlation operator $\op{C}$ that acts on uncorrelated
product states $\ket{\Psi}$
\begin{equation}
  \ket{\hat{\Psi}} = \op{C}\ \ket{\Psi}
\end{equation}
such that the many-body state $\ket{\hat{\Psi}}$ contains the short
ranged correlations.
For the correlator we make the following ansatz
\begin{equation} \label{eq:ComCr}
  \opC = \opCom \cdot \opCr =\exp\Bigl\{-\I \sum_{i<j}\op{g}_{\Omega\,ij}\Bigr\}
                \cdot \exp\Bigl\{-\I \sum_{i<j}\op{g}_{r\,ij}\Bigr\} \ .
\end{equation}
It is the product of a radial correlator $\opCr$ and a tensor
correlator $\opCom$, both, expressed with a hermitian two-body generator in
the exponent.

\subsection{Cluster expansion}

As the ansatz for the correlator contains a two-body operator in the
exponent any correlated operator will contain many-body parts. For
example a Hamiltonian consisting of one- and two-body parts will turn
into
\begin{eqnarray}
  \op{\hat{H}}&=&\op{C}^\dagger \op{H} \op{C}=
  \op{C}^\dagger \Bigl( \sum_i T_i + \sum_{i<j} V_{ij} \Bigr)\op{C}\\
  \nonumber
  &=&\sum_i T_i +  \sum_{i<j} \hat{T}_{ij}^{[2]} + \sum_{i<j<k}
         \hat{T}_{ijk}^{[3]} + \cdots +\sum_{i<j} \hat{V}_{ij}^{[2]} +
         \sum_{i<j<k} \hat{V}_{ijk}^{[3]} + \cdots \ ,
\end{eqnarray}
where the upper script $^{[n]}$ indicates irreducible n-body
operators.  Here we introduce an approximation by keeping terms only
up to two-body operators. This approximation should be good for
systems where the range of the correlator ($g_{r\,ij}=0$ and
$g_{\Omega\,ij}=0$ for $r_{ij}>R_{c}$) is short compared to the mean
particle distances. In that case the probability to find 3 particles
simultaneously within the correlation range $R_{c}$ is small.

\subsection{Radial correlator}

The radial correlator $\opCr$ (described in detail in \cite{ucom98})
shifts a pair of particles in the radial direction away from each
other so that they get out of the range of the repulsive core.  To
perform the radial shifts the generator of the radial correlator uses
the radial momentum operator $\oppr$ together with a shift function
$s(r)$ that depends on the distance of the two nucleons.
\begin{equation}
  \op{g}_{r\,ij}=\frac{1}{2}
  \Bigl( \oppr{}_{ij} s(r_{ij}) + s(r_{ij}) \oppr{}_{ij} \Bigr)
  \label{eq:Cr}
\end{equation}
The shift function $s(r)$ is optimized to the potential under
consideration.  It is large for short distances and will vanish at
large distances.

\begin{figure}[tb]
  \begin{center}
    \includegraphics[angle=0,width=0.9\linewidth]{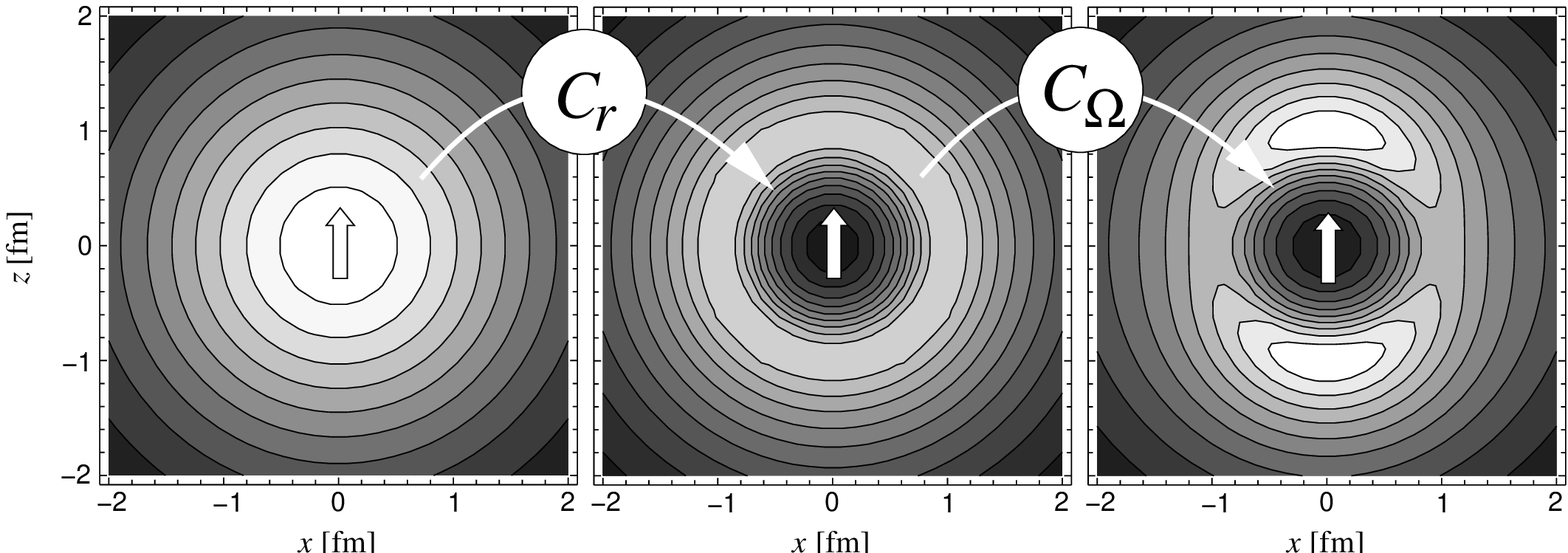}\\[5mm]
    \includegraphics[angle=0,width=0.7\linewidth]{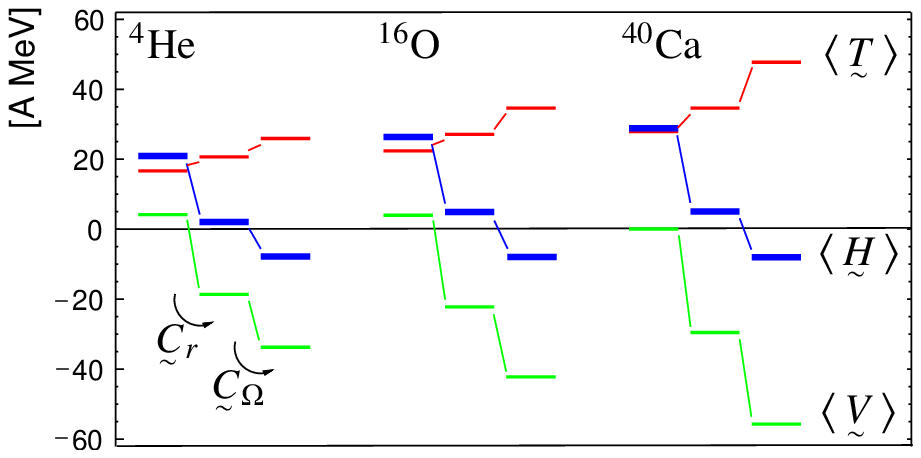}
  \end{center}
  \vspace{-2mm}
  \caption{Upper part: Two-body density
    $\rho^{(2)}_{S,T}(\vec{r}_1\!\!-\!\vec{r}_2)$ of $^4$He for a pair
    of nucleons with isospin $T\!=\!0$ and parallel spins, $S\!=\!
    M_S\!=\!1$. Bright ares denote large probabilities.
    Arrow indicates spin direction and
    $(x,y,z)=(\vec{r}_1\!\!-\!\vec{r}_2)$ relative distance vector.
    Lower part: corresponding kinetic, potential and total energies
    per particle of $^4$He, $^{16}$O and $^{40}$Ca, without, with
    radial, and with radial and tensor correlations (Bonn-A
    potential).}
  \label{fig:BonnA-Energies}
\end{figure}

The effect of the transformation $\ket{\Psi}\rightarrow
\op{C}_r\ket{\Psi}$ is shown in the upper part of Fig.
\ref{fig:BonnA-Energies} where the two-body density $\rho^{(2)}_{S,T}$
is displayed as a function of the distance vector
$(\vec{r}_1\!-\!\vec{r}_2)$ between two nucleons in $^4$He.  On the
l.h.s. $\rho^{(2)}_{S,T}$ is calculated with the shell-model state
$\ket{(1s_{1/2})^4}$ that is just a product of 4 Gaussians.  It has a
maximum at zero distance which is in contradiction to the short ranged
repulsion of the interaction.  This inconsistency is removed by
application of the radial correlator $\op{C}_r$ that moves density out
of the region where the potential is repulsive.  The corresponding
kinetic, potential and total energies are displayed in the lower part
of the figure for three nuclei.  The radially correlated kinetic
energy $\expect{\op{C}_r^\dagger\op{T}\op{C}_r}$ increases somewhat
compared to $\expect{\op{T}}$ but this is overcompensated by the gain
of about -25~MeV per particle in the correlated potential energy.
Nevertheless the nuclei are still unbound.

\subsection{Tensor correlator}

The tensor force in the $S\!=\!1$ channels of the nuclear interaction
depends on the spins and the spatial orientation
$\unitvec{r}=(\vec{r}_1\!-\!\vec{r}_2)/(|\vec{r}_1\!-\!\vec{r}_2|)$ of the
nucleons via the tensor operator
\begin{equation}
  S_{\!12}(\unitvec{r},\unitvec{r}) =
  3 (\vec{\sigma}_1\cdot\unitvec{r})(\vec{\sigma}_2\cdot\unitvec{r}) -
  (\vec{\sigma}_1\cdot\vec{\sigma}_2) =
  2\ \Bigl( 3 (\vec{S}\cdot\unitvec{r})^2 - \vec{S}^2 \Bigr) \eqdot
\end{equation}
An alignment of $\unitvec{r}$ with the direction of the total spin
$\vec{S}= \frac{1}{2}(\vec{\sigma}_1\!+\!\vec{\sigma}_2)$ is favored
energetically.  The tensor correlator $\opCom =\exp\Bigl\{-\I
\sum_{i<j}\op{g}_{\Omega\,ij}\Bigr\}$, defined by the generator
\begin{equation}
  \op{g}_{\Omega\,ij} = \vartheta(r_{ij})
  \frac{3}{2}\Bigl(
  (\vec{\sigma}_i \vec{p}_{\Omega\,ij})(\vec{\sigma}_j \vecr_{ij}) +
  (\vec{\sigma}_i \vecr_{ij})(\vec{\sigma}_j \vec{p}_{\Omega\,ij})
  \Bigr)  \ ,
  \label{eq:Com}
\end{equation}
achieves this alignment by shifts perpendicular to the relative
orientation $\unitvec{r}_{ij}$.  For that the generator of the tensor
correlator uses a tensor operator constructed with the orbital part of
the relative momentum operator $\vec{\op{p}}_{\Omega\,ij} = \vecp_{ij}
\!-\! \vec{\op{p}}_{r\,ij}$.  The $r$-dependent strength and the range of
the tensor correlations is controlled by $\vartheta(r)$.  For details
see \cite{ucom03}.

The application of the tensor correlator $\op{C}_\Omega$ leads to the
two-body density depicted in the right hand contour plot of
Fig.~\ref{fig:BonnA-Energies}.  One may visualize the action of
$\op{C}_\Omega$ as a displacement of probability density from the
`equator' to both `poles', where the spin of the $S\!=\!1$ component
of the nucleon pair defines the `south-north' direction.  Again this
costs kinetic energy but now the many-body state is in accord with the
tensor interaction and one gains the binding needed to end up with
about -8~MeV per particle.

\section{Interaction in momentum-space}

\begin{figure}[tb]
  \includegraphics[angle=0,width=\textwidth]{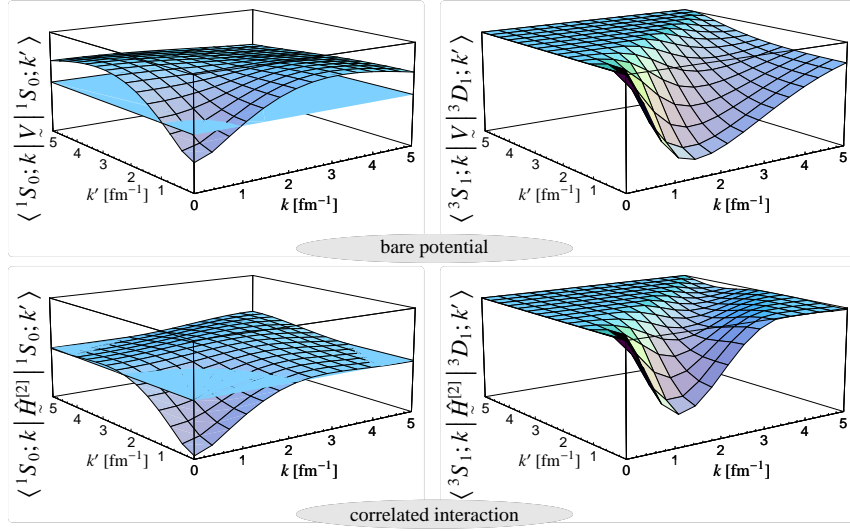}
  \caption{Bare and correlated Argonne V8' interaction in
    momentum-space. Matrix elements range from -2 to +2 $\fm^{-1}$ on
    the l.h.s. and from -1.5 to 0 $\fm^{-1}$ on the r.h.s.}
  \label{fig:vlowk}
\end{figure}

The inclusion of the short-range correlations achieves a
\emph{pre-diagonalization} of the nuclear hamiltonian that is
illustrated in Fig.~\ref{fig:vlowk}. On the l.h.s. the effect of the
radial correlations is shown in the ${}^1S_0$ channel. The
correlated interaction evaluated in momentum-space is more attractive
and does not possess the large off-diagonal matrix elements of the
bare interaction. Also the tensor components of the correlated
Hamiltonian do not connect to high momenta as is illustrated with the
matrix elements between the ${}^3S_1$ and the ${}^3D_1$ channel. The
correlated interaction is therefore a low-momentum interaction very
similar to the $V_{\mathit{low-k}}$ \cite{vlowk03}.

\section{Effective interaction}

To test the two-body approximation we performed no-core shell model
calculations with the correlated AV8' interaction for $^4$He and
compared with exact results \cite{hirschegg03,refav8p}.  It turned out
that neglecting the $n=$ 3- and 4-body parts $\hat{\op{T}}^{[n]}$ and
$\hat{\op{V}}^{[n]}$ of the correlated Hamiltonian leads to an
overbinding which is of the same order as the contribution from
genuine 3-body forces.  Because of the low-momentum nature of the
correlated Hamiltonian
$\hat{\op{H}}^{C2}\!=\!\op{T}\!+\!\hat{\op{T}}^{[2]}\!+\!\hat{\op{V}}^{[2]}$
it can be used directly with simple model spaces built of Slater
determinants. The effects of missing higher-order contributions to the
correlated interaction and of genuine three-body forces is for now
effectively described by a momentum-dependent central and spin-orbit
correction $\op{H}_{corr}$ with four parameters adjusted to four
doubly magic nuclei.  The resulting effective interaction
$\hat{\op{H}}_{e\!f\!f}\!=\!\hat{\op{H}}^{C2}+\op{H}_{corr}$ is used
for all nuclei up to mass number 60.  The expectation value of
$\op{H}_{corr}$ is typically 15\% of the correlated interaction
energy.

\section{Hartree-Fock calculations}

\begin{figure}[t]
  \includegraphics[angle=0,width=\textwidth]{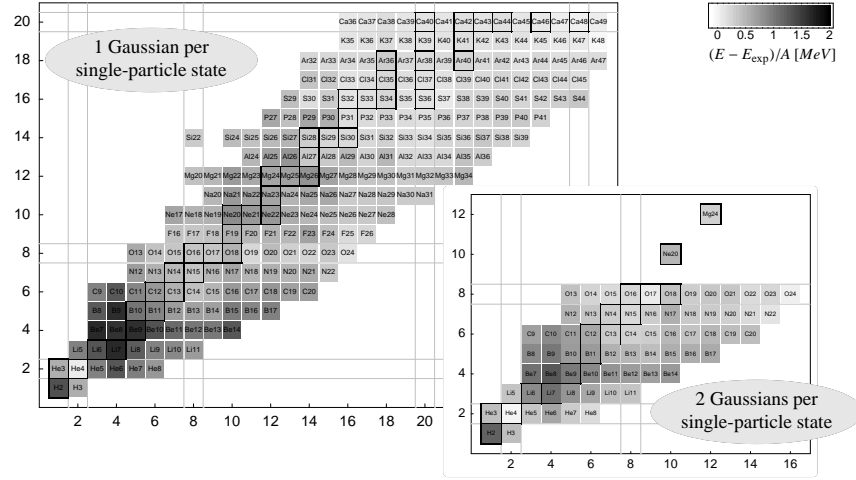}
  \caption{Deviation of mean-field binding energies from measured ones.}
  \label{fig:NuclChart}
\end{figure}

\begin{figure}[tb]
  \begin{center}
   \includegraphics[angle=0,width=0.31\textwidth]{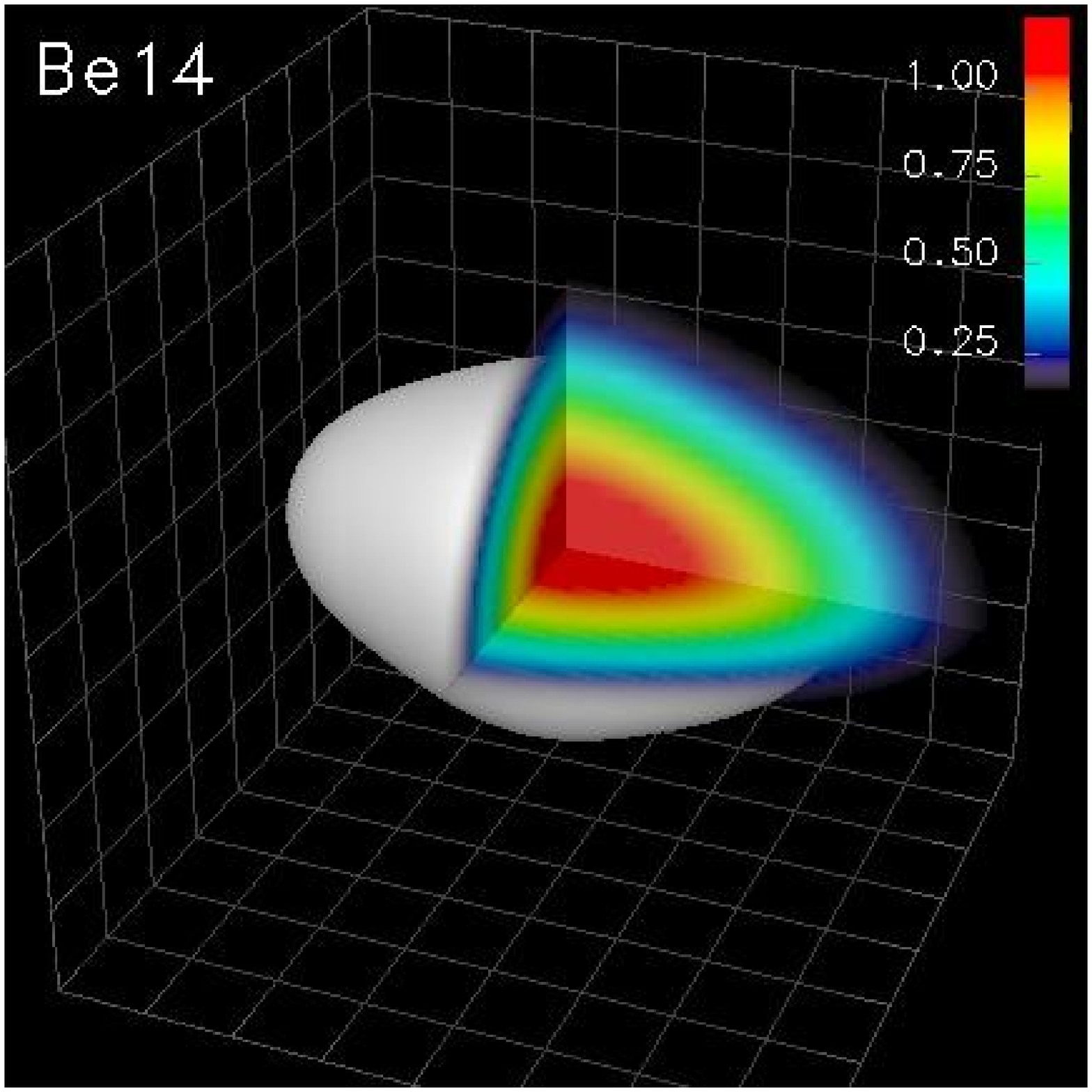}
   \includegraphics[angle=0,width=.31\textwidth]{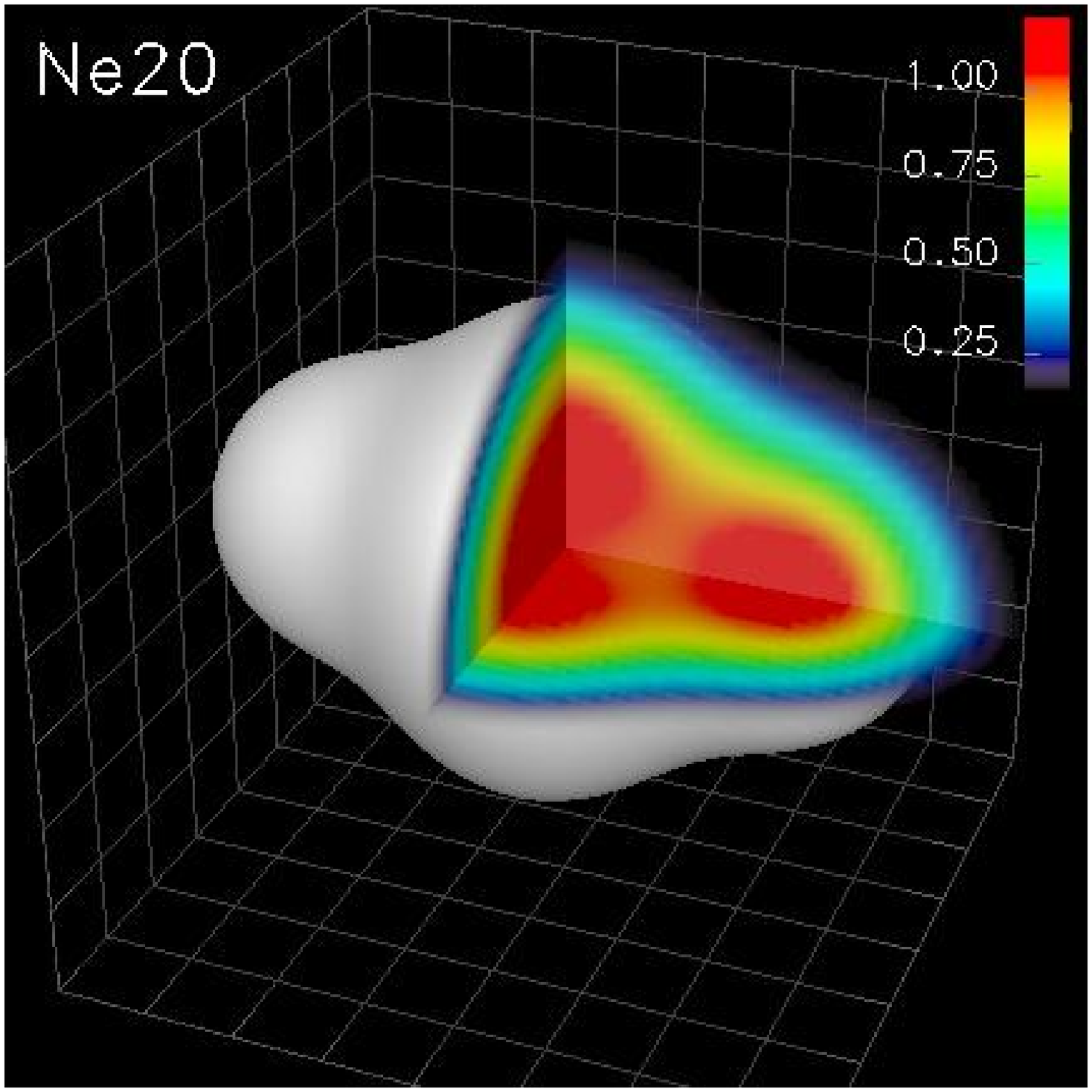}
   \includegraphics[angle=0,width=.31\textwidth]{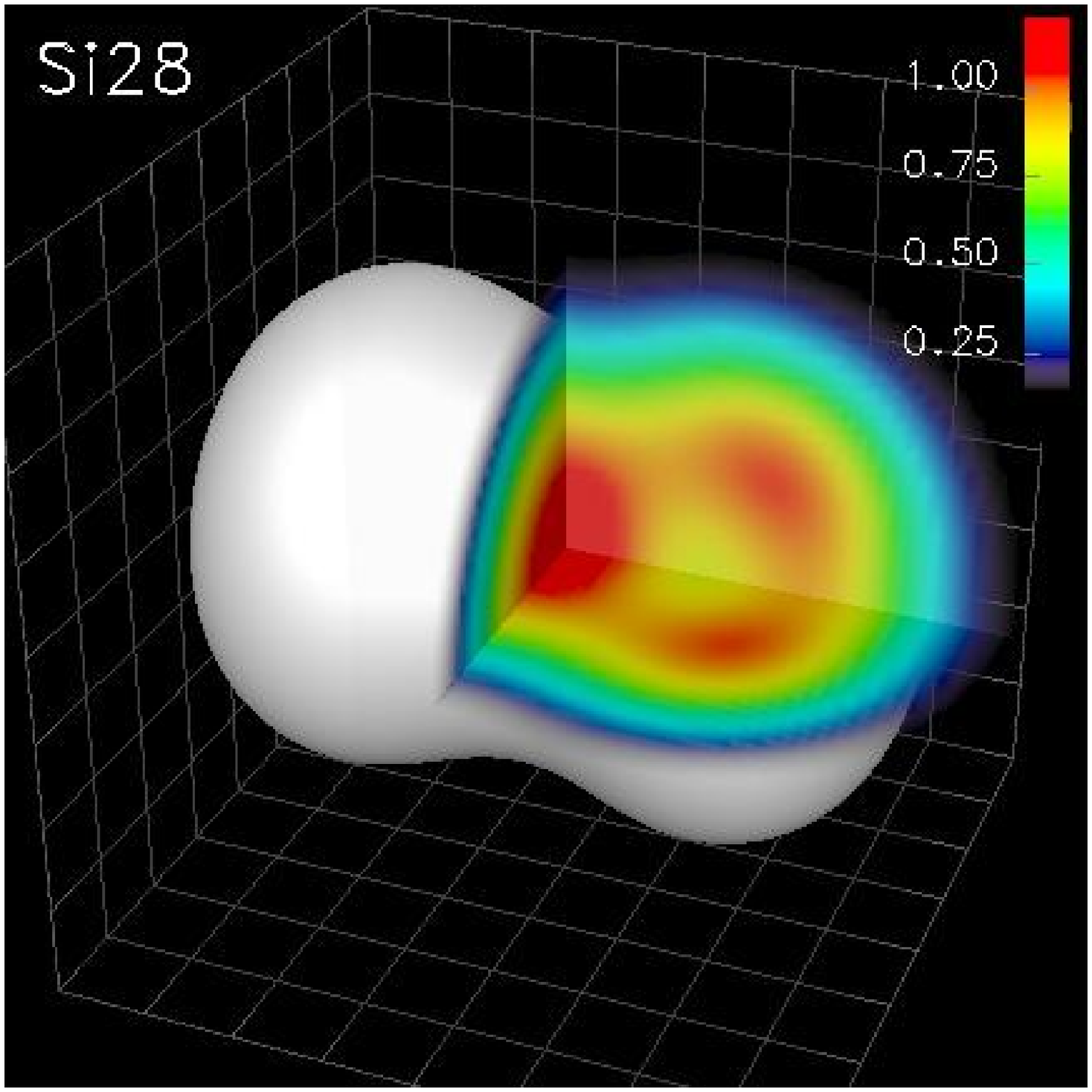}\\
   \includegraphics[angle=0,width=.31\textwidth]{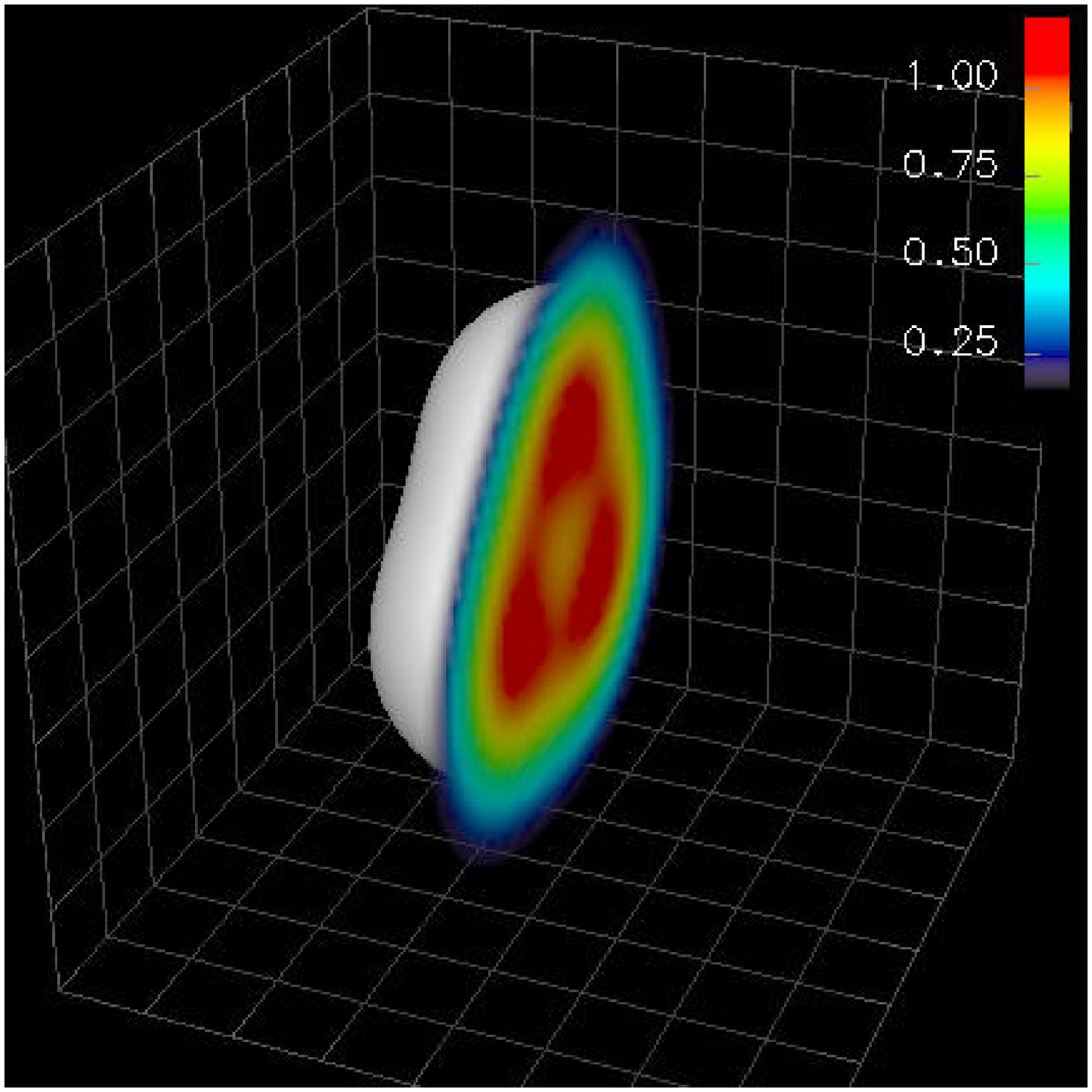}
   \includegraphics[angle=0,width=.31\textwidth]{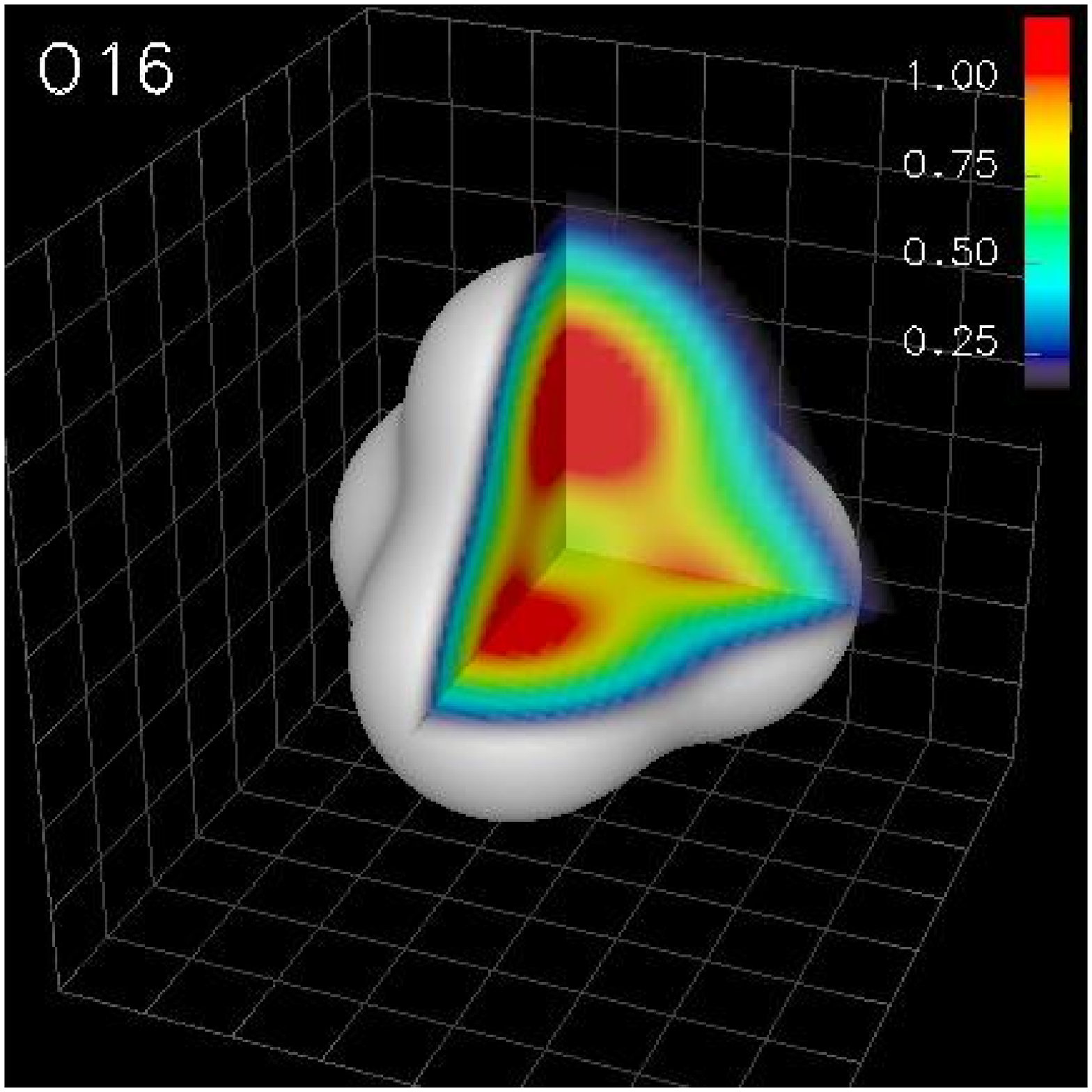}
   \includegraphics[angle=0,width=.31\textwidth]{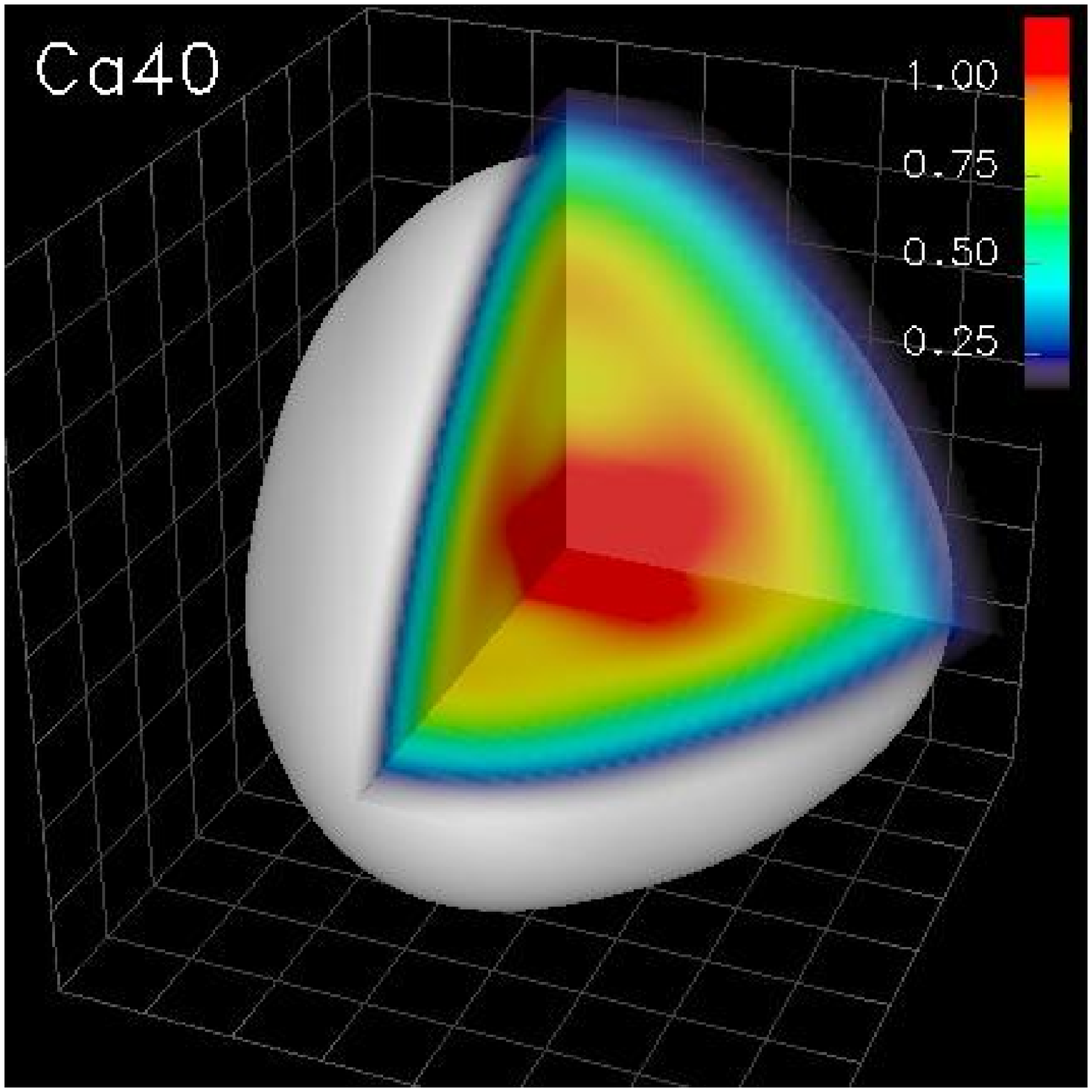}
 \end{center}
 \caption{One-body densities of intrinsic states.
   Upper part: PAV calculation $^{14}$Be, $^{20}$Ne, $^{28}$Si (local minimum);
   lower part: VAP calculation $^{12}$C,  $^{16}$O, $^{40}$Ca.} 
 \label{fig:SelectedShapes}
\end{figure}

For the nuclei listed in the nuclear chart Fig.~\ref{fig:NuclChart} we
minimized the expectation value
$\matrixe{\Psi}{\hat{\op{H}}_{e\!f\!f}-\op{T_{cm}}}{\Psi}/\braket{\Psi}{\Psi}$
with respect to all parameters of the single-particle states
\begin{equation}
\braket{\vec{x}}{q}=\sum_{i} c_i \exp \biggl\{ -
       \frac{(\vec{x} -\vec{b}_i)^2}{2a_i} \biggr\}
        \ket{\chi_i} \otimes \ket{\xi}
\end{equation}
that are contained in the Fermionic Molecular Dynamics \cite{fmd00}
state $\ket{\Psi}$ (Eq.~(\ref{Slater})). The summation is either for
one or for two independent Gaussians per single-particle state.

The inclusion of a second Gaussian improves the masses of $p$-shell
nuclei substantially (inset of Fig.~\ref{fig:NuclChart}).  The largest
deviations occur for nuclei with an $\alpha$-cluster structure like
$^8$Be or $^{12}$C or for intrinsically deformed nuclei at the middle
of the $sd$-shell.  Some examples of intrinsic shapes are displayed in
the upper half of Fig.~\ref{fig:SelectedShapes}. In $^{14}$Be the
extra six neutrons have pulled together the well localized pair of
$\alpha$'s that form $^8$Be. The peanut like shape for $^{28}$Si is a
local minimum.

\section{Projection after variation and variation after projection}

To improve the many-body Hilbert space we project on spin and parity
after variation (PAV). We also perform variation after projection
(VAP) calculations in the sense of the generator coordinate method.
The intrinsic state is minimized here with constraints on radius,
dipole moment, quadrupole or octupole moment.

With the above described effective interaction $^{16}$O gains in a VAP
calculation about 5~MeV in binding by forming $\alpha$-clusters (see
lower part of Fig.~\ref{fig:SelectedShapes}) compared to the spherical
closed-shell configuration which represents the energy minimum in the
PAV case.  Even for $^{40}$Ca VAP leads to a deformation towards a
tetrahedron of ten $\alpha$'s which however are much more amalgamated
than in $^{16}$O (Fig.~\ref{fig:SelectedShapes}).

\begin{figure}[tb]
  \setlength{\unitlength}{\textwidth}
  \begin{picture}(1,0.6)
    \put(0,0.3){\makebox(1,0.3){
        \includegraphics[angle=0,width=.25\unitlength]{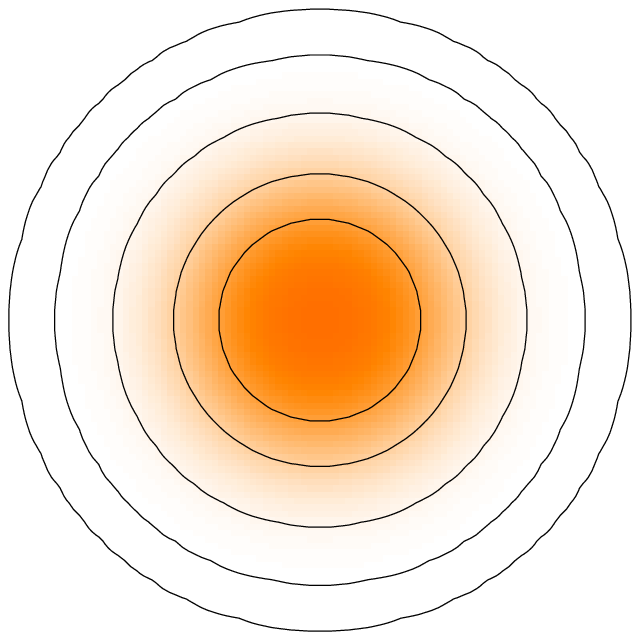}
        \includegraphics[angle=0,width=.25\unitlength]{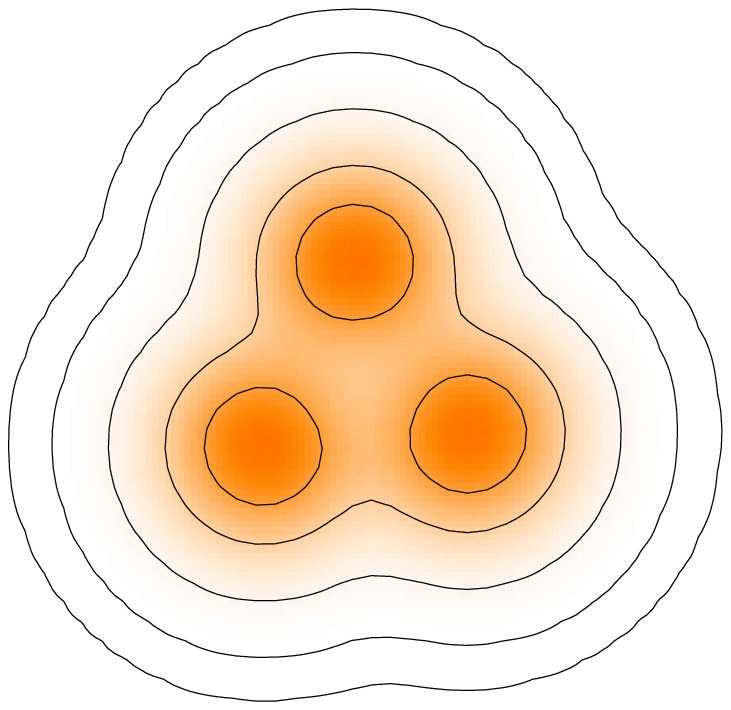}
        \includegraphics[angle=0,width=.25\unitlength]{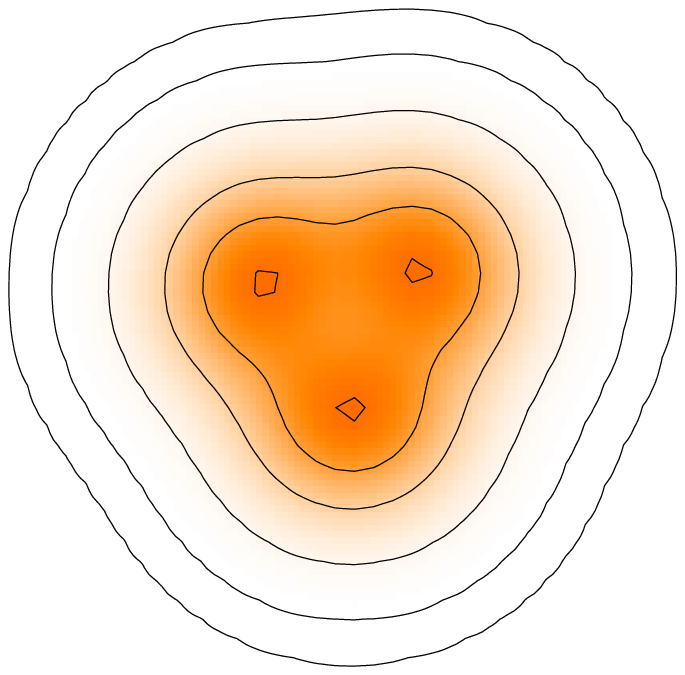}
      } }
    \put(0.2,0.31){V/PAV} \put(0.4,0.31){VAP $\alpha$-cluster} \put(0.73,0.31){VAP}
    \put(0,0.0){\makebox(1,0.3){
        \includegraphics[angle=0,width=.25\unitlength]{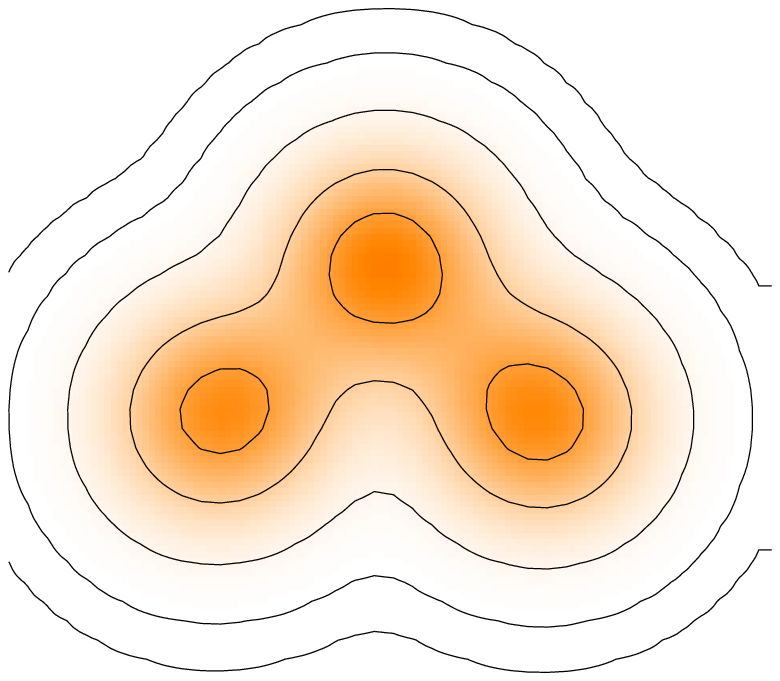}
        \includegraphics[angle=0,width=.25\unitlength]{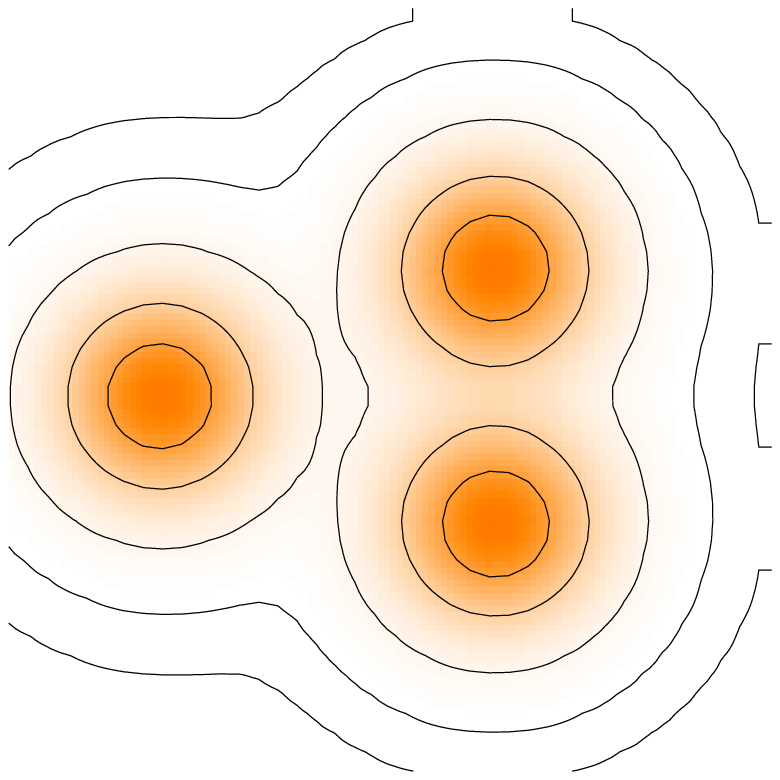}
        \includegraphics[angle=0,width=.25\unitlength]{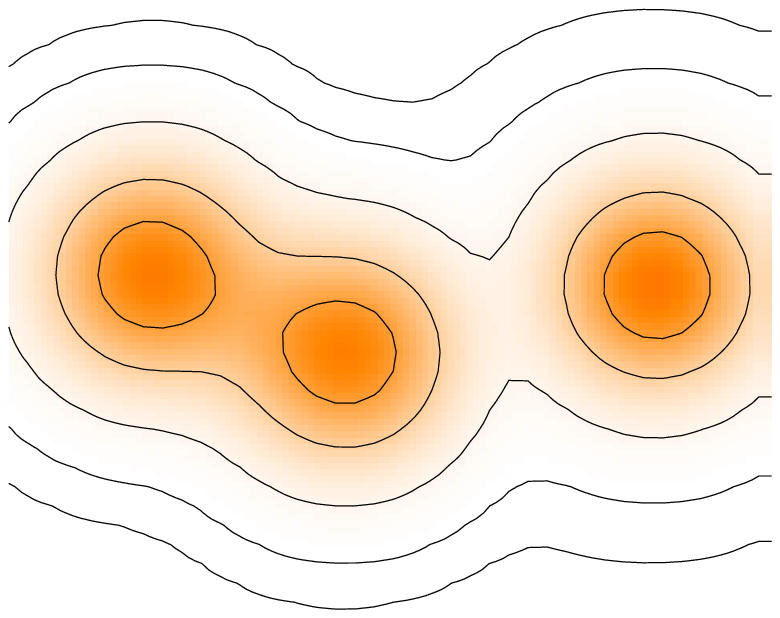}
      } }
    \put(0.21,0.0){`` $3_1^-$ ''} \put(0.49,0.0){`` $0_2^+$ ''}
    \put(0.74,0.0){`` $0_3^+$ ''}
  \end{picture}
  \caption{One-body densities of intrinsic $^{12}$C states.}
  \label{fig:C12Shapes}
\end{figure}

In a next step the different intrinsic shapes obtained in the VAP
process can be used to perform multiconfiguration calculations.
\vspace{1ex}

A very interesting nucleus is $^{12}$C for which the shell-model
configuration $(1s_{1/2})^4 (1p_{3/2})^8$ is competing with the
3-$\alpha$ cluster structure \cite{enyo98,itagaki03}. Like in $^{16}$O
the PAV ground state of $^{12}$C turns out to be a spherical
shell-model state (Fig.~\ref{fig:C12Shapes}, V/PAV).  However,
variation after projection leads to a triangular shape made of three
$\alpha$'s (Fig.~\ref{fig:C12Shapes}, VAP) and 7.2~MeV of additional
binding as indicated in Table~\ref{tab:c12}. A pure $\alpha$-cluster
configuration obtained in a VAP calculation has larger distances
between the $\alpha$-clusters and is 4.3~MeV less bound than the
shell-model configuration.  The difference to the full VAP is due to
the polarization of the $\alpha$-clusters. The description of this
polarization is significantly improved by using two Gaussians per
single-particle state.

\begin{table}[b]
  \small
  \centering
  \begin{tabular}{@{}lrrr@{}} \hline 
        & $E_b \; [\MeV]$ & $r_{\mathit{charge}}\; [\fm]$ & $B(E2) \;
        [e^2 \fm^4]$ \\ \hline
        V/PAV & 84.7 & 2.33 & - \\
        VAP $\alpha$-cluster & 80.4 & 2.66 & 56.3 \\
        VAP & 91.9 & 2.38 & 24.7 \\
        Multiconfig & 93.4 & 2.50 & 40.0 \\ \hline
        Exp & 92.2 & 2.47 & $39.7 \pm 3.3$\\ \hline
  \end{tabular}
  \caption{Binding energies, charge radii and $BE(2)$-values of $^{12}$C.}
\label{tab:c12}
\end{table}

The description of the excited states requires an enlarged Hilbert
space.  Multiconfiguration calculations with the configurations shown
in the lower part of Fig.~\ref{fig:C12Shapes} show only a small
increase in binding for the groundstate but have a significant effect
on the radius and the $B(E2)$ value for the $0_1^+ \longrightarrow
2_1^+$ transition. The additional configurations have been chosen to
give lowest energies for the $3_1^-$ and the second and third $0^+$
states. We find 3 $\alpha$-cluster structures for the $0_2^+$ state
and ${}^8$Be + $\alpha$-cluster structures for the $0_3^+$ state. For
an improved description of these states a larger number of
configurations with greater distances between the $\alpha$'s is
needed. This is consistent with an assumed Bose condensed
state\cite{horiuchi03} for the $0_2^+$ state.

\begin{figure}[tb]
  \centering
  \includegraphics[angle=90,width=0.65\textwidth]{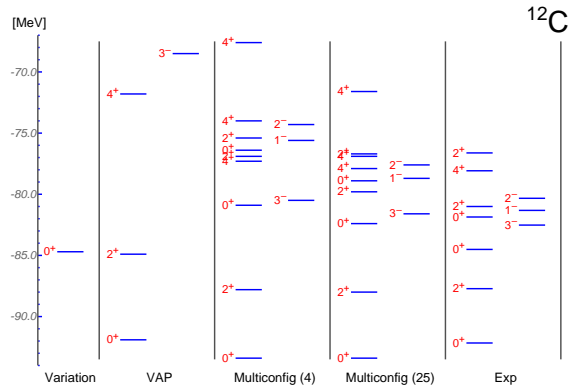}
  \caption{Calculated and experimental level scheme for ${}^{12}$C.}
  \label{fig:c12levelscheme}
\end{figure}


\end{document}